# Non-Universal and Universal Aspects of The Large Scattering Length Limit

Gerald A. Miller

*Department of Physics, University of Washington, Seattle, WA 98195-1560*

(Dated: October 8, 2018)

The momentum density, $n(k)$ of interacting many-body Fermionic systems is studied (for $k > k_F$) using examples of several well-known two-body interaction models. It is shown that $n(k)$ can be approximated by a zero-range model for momenta $k$ less than about $0.1/r_e$, where $a$ is the scattering length and $r_e$ the effective range. If the scattering length is large and one includes the effects of a fixed value of $r_e \neq 0$, $n(k)$ is almost universal for momenta $k$ up to about $2/r_e$. However, $n(k)$ can not be approximated by a zero-range model for momenta $k$ greater than about $1/(ar_e^2)^{1/3}$, if one wishes to maintain a sum rule that relates the energy of a two component Fermi-gas to an integral involving the density. We also show that the short separation distance, $s$, behavior of the pair density varies as $s^6$.

Interacting many-body Fermionic systems are copious in nature, with examples occurring in astrophysics, nuclear physics, condensed matter physics, and most recently in atomic systems. The development of trapping, cooling and magnetic resonance techniques techniques for ultracold atoms allows the strength of the two-body interaction to be controlled experimentally [1–3]. At low relative energies this strength is characterized by the scattering length, $a$, which can be much, much larger than the range of the two-body interactions, $R$.

Such systems are of interest to understanding the nucleon-nucleon interaction, which is characterized at low energies by scattering lengths of magnitude much larger than the effective range. Indeed, the limit of $a \to \infty$, $R \to 0$, characterized as the unitary limit, has been used as a benchmark problem for nuclear many-body physics. See for example, G.F Bertsch in.Ref. [4] and for example [5, 6].

If $a/R$ approaches infinity, the system is expected to have universal properties that are determined only by the scattering length. Tan [7–9] derived universal relations between diverse properties of any arbitrary system consisting of fermions in two spin states with a large scattering length. These relations include the coefficient of the $1/k^4$ tail of the momentum distribution [7], a decomposition of the energy into terms that insensitive to short distances [7], an expression for the local pair density [7], and various other properties of interacting many fermion systems [8–12]. Tan's derivations start with the assumption that the interaction between constituents is a zero-range pseudopotential. Brataan and Platter [13] confirmed these relations using a zero-range interaction, renormalized by cutting off the intermediate momentum integrals at high values of the momentum. But the range of interaction is never 0 for any physical system even though the ratio $R/a$ can be made exceedingly.

The aim of the present epistle is to explore the consequences of the non-zero range of interaction. We study how the effects of the non-zero value of the effective range influence the relative two-fermion wave function of the interacting two-bodies. The square of the momentum-space wave function, $\widetilde{\phi}(k)$ determines the shape of the system's momentum distribution, $n(k) = \widetilde{\phi}^2(k)$ for $k > k_F$, the Fermi momentum. Recent studies that examine the non-zero nature of the effective range include [14–16]. Ours focuses on exhibiting the most relevant features of the two-body system

We note that the universal relations all involve a property of the system that depends on the contact density, $\mathcal{C}(\mathbf{R})$, which is the local density of interacting pairs. Its integral over volume is denoted the contact, $C$ [7], said to be a measure of the number of atom pairs with large (but not too large) relative momentum [17]. Recent works that apply the contact formalism to nuclear physics include Refs. [18–21].

At low relative energies the s-wave scattering phase shift $\delta$ can be expressed in terms of the effective range expansion:

$$k \cot \delta = -\frac{1}{a} + \frac{1}{2} r_e k^2 + \cdots, \qquad (1)$$

where $k$ is the relative momentum ($\hbar^2/m$ is taken as unity by convention), $a$ is the scattering length and $r_e$ is the effective range. The effective range is expected to be of the order of range of the two-body interactions that govern the system, but $a$ can be much larger in magnitude. The present analysis is concerned with cases for which $a > 0$. If $a \gg r_e > 0$, the S-matrix element $e^{2i\delta(k)}$ has a pole on the positive imaginary axis. This pole corresponds to the energy of a bound state of very small binding energy, $B = 1/a^2 + \frac{r_e}{a^3} + \cdots$, in units with $\hbar^2/m = 1$, with $m$ as twice the reduced mass of the interacting pair.

With a zero range interaction, the resulting wave function is simply $\phi_0(r) = u_0(r)/r = \sqrt{2/(4\pi a)} \exp(-r/a)/r$, with the momentum space version, $\widetilde{\phi}_0(k) = \sqrt{8\pi a}\, a/(k^2 a^2 + 1)$, and $n_0 = \widetilde{\phi}_0^2$. This function is the source of the claimed $1/k^4$ behavior of the density. The range of validity of these expressions is said [7, 17] to be

$$1/a \ll k \ll 1/r_e. \qquad (2)$$

If $r_e$ is taken to 0, then the upper limit would be infinite.

But $r_e \neq 0$ for all physical systems, so that other momentum scales may enter. For example, consider the effective range expansion of Eq. (1). In the large $a$ limit, the first term is very small. Thus the second term can be as large as or much larger than the second term for relatively small values of $k$. For example, if $k = \sqrt{\frac{2}{a r_e}}$, the second term of Eq. (1) provides a 100% correction to the first term. An effect of that size cannot be ignored. The calculations will show that other momentum scales smaller than $1/r_e$ are important.

We study the influence of the non-zero range of the interaction as manifest by the difference between $n(k)$ and $n_0(k)$. The physics of the interior must matter because, $(\nabla^2 + \frac{1}{a^2}) e^{-r/a}/r = -4\pi \delta(\mathbf{r})$, so that the function

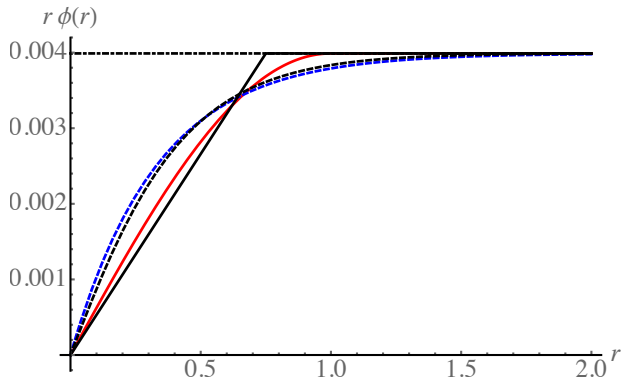

FIG. 1. (Color online) Comparison of $r\phi_S(r)$ (red) solid, $r\phi_H(r)$ (blue) dashed, and $r\phi_0(r)$ (black) dot-dashed, $r\phi_{SD}(r)$ (black) solid, and $r\phi_E(r)$ (black) dashed for $a/R = 10,000$. Dimensionless units are used (see text).

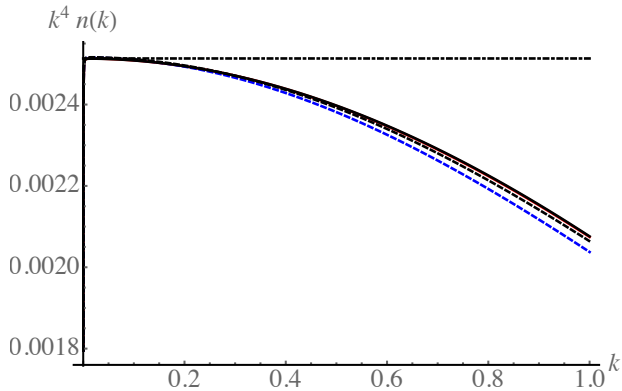

FIG. 2. (Color online) Comparison of $k^4 n(k)$, for five different wave functions with $n(k) \equiv \widetilde{\phi}^2(k)$. Square well)(red) solid, Hulthein (blue) dashed, zero-range (black) dot-dashed, surface delta (black) solid, and exponential (black)

$e^{-r/a}/r$ is not a solution of the Schroedinger equation in the usual sense. The region with $r \leq r_e$ matters, $r_e$ cannot be set to 0, and the range of validity of the $1/k^4$ behavior must be limited.

Our approach is to analyze four simple available models of the interaction that have the same non-zero effective range, and then to draw some general conclusions. Starting with an attractive, square-well, ($S$), potential of depth $V_0$ and range $R$ is useful. The bound state wave function is given by

$$\phi_S(r) = \frac{N_S}{r}[\frac{\sin Kr}{\sin KR}\theta(R-r) + e^{-(r-R)/a}\theta(r-R)], \quad (3)$$

where $N_S$ is a normalization factor, and $K = \sqrt{V_0 - B}$. For very large scattering lengths $KR$ is slightly larger than $\pi/2$, and the effective range $r_e$ is very close to $R$, the range of the interaction. The momentum space wave function is the Fourier transform:

$$\widetilde{\phi}_S(k) = \bar{N}_S \frac{\left(\frac{\sin(kR)}{ak} + \cos(kR)\right)}{(K^2 - k^2)(1 + a^2 k^2)}. \quad (4)$$

Another simple model is the Hulthein wave function, ($H$). The bound state wave function is given by

$$\phi_H(r) = \frac{\sqrt{\beta(\beta+\alpha)}}{\beta - \alpha}\sqrt{\frac{2\alpha}{4\pi}}\frac{1}{r}(e^{-\alpha r} - e^{-\beta r}), \quad (5)$$

with $\beta \gg a$, $B = \alpha^2$. In the large $a/r_e$ limit, $\alpha = 1/a$. The momentum space wave function is given by

$$\widetilde{\phi}_H(k) = 2\sqrt{2\pi}\sqrt{\alpha\beta(\alpha+\beta)^3}\frac{1}{\alpha^2+k^2}\frac{1}{\beta^2+k^2}. \tag{6}$$

The surface delta ($SD$) function potential $V(r) \propto \delta(r-R)$ is also easily analyzed. The coordinate space wave function is given by

$$\phi_{SD}(r) = \frac{N_{SD}}{r}\left(\frac{\sinh(r/a)}{\sinh(R/a)}\theta(R-r) + e^{-(r-R)/a}\theta(r-R)\right), \tag{7}$$

if $B = 1/a^2$. The momentum space wave function is given by

$$\widetilde{\phi}_{SD}(k) = \tilde{N}_{SD}\frac{1}{1+(ka)^2}\frac{\sin(kR)}{kR}. \tag{8}$$

The exponential potential ($E$), $V(r) = -V_{0E}e^{-\mu r}$ provides another well-studied example. A bound state wave function is given by

$$r\phi_E(r) = N_E J_{2\sigma/\mu}(2\sqrt{V_{0E}}/\mu\, e^{-\mu r/2}), \tag{9}$$

with $V_0 > 0$, and the value of $\sigma$ determined by the condition $J_{2\sigma/\mu}(2\sqrt{2V_{0E}/\mu}) = 0$, $B = -\sigma^2$, and $1/\sigma = a$ in the large scattering length limit. A useful form of the momentum-space wave function is obtained by using the power-series expansion for the Bessel function. The large scattering length result is given by

$$\widetilde{\phi}_E(k) = \tilde{N}_E \sum_{j=0}^{\infty}(\frac{V_{0E}}{\mu^2})^{j+1/(a\mu)} \times$$
$$\frac{(-1)^j}{j!\Gamma(j+1+2/(\mu a))}\frac{1}{k^2+(1/a+\mu j)^2}. \tag{10}$$

This function can be seen as a generalization of the Hulthein wave function.

Five coordinate space wave functions ($0, S, H, SD, E$) are compared in Fig. 1. The examples shown here use the very large ratio for $a/R = 10000$ to ensure that any non-universal features do not arise from an insufficiently large scattering length. The values $V_{0E} = 18.73886$, $\mu = 3.6$ are used to obtain same values of $a, r_e$ as for the other potentials. The effective range is chosen to be unity in the natural length unit of the system. If the binding energy is very small, the effective range can be computed using the bound state wave function. The relevant expression is (for $B = -1/a^2$)

$$r_e = 2\int dr(e^{-2r/a} - u_{ER}^2(r)), \tag{11}$$

where $u_{ER}(r) = r\phi(r)$ normalized so that its asymptotic form is $e^{-r/a}$. This expression is taken from the usual effective range expansion [22], but using $k = i/a$: the energy is taken to approach 0 from negative values. Any differences from the approach using positive energy are accounted for by higher order terms in the effective range expansion, which are significant for small values of $k$. For the square well, $R \approx 1$. For the Hulthein wave function one finds:

$$r_e = 2\int_0^\infty dr(e^{-2r/a} - (e^{-r/a} - e^{-\beta r})^2) \approx \frac{3}{\beta}, \tag{12}$$

under the assumption that $\beta a \gg 1$. For the surface delta model, the $\delta(r-R_{SD})$ occurs at $R_{SD} = 3/4 r_e$. Fig. 1 shows that the effects of the finite range of interaction are considerable for $r < 1$. Using any value of $a/R$ greater than about 400 would lead to very similar coordinate-space plots.

Five momentum space wave functions are compared (in terms of $n(k) = \widetilde{\phi}^2(k)$) in Fig. 2. The zero range model shows the $1/k^4$ behavior, and all of the models look like the zero range model for values of $k$ less than about 0.1. Thus Eq. (2) is verified, if one takes the symbol $\ll$ to mean about 0.1. For larger values there is a small, but distinct, fall-off behaving quadratically in $k$ that is essentially the same for each model. We show below that this seemingly innocuous effect has significant consequences for the energy sum rule of [7],

First it is necessary to improve the understanding of $n(k)$. Consider the Lippmann-Schwinger equation:

$$\widetilde{\phi}(k) = -\frac{4\pi}{k^2+B} \int dr \, \frac{\sin(kr)}{k} V(r) r\phi(r) = -\frac{4\pi}{k^2+B} T(k),$$
$$T(k) \equiv \int dr \, \frac{\sin(kr)}{k} V(r) r\phi(r). \tag{13}$$

The quantity $T(k)$ is an off-shell transition matrix element, and its integrand is non-vanishing only for values of $r$ within the range of the two-body potential, $V(r)$. Repeated integration by parts in Eq. (13) shows that $\widetilde{\phi}$ falls as $1/k^4$ for asymptotic values of $k$, provided $V(r)r\phi$ does not vanish at the origin. The falloff is at least as fast as $\sim 1/k^6$ if $V(r)r\phi$ does vanish at the origin. If the potential is a function of $r^2$, $\widetilde{\phi}(k)$ falls off faster than any inverse power of $k^2$. This means that the statement e.g. [23] that $C = \lim_{k\to\infty} k^4 n(k)$ cannot be correct because the correct value of the limit is 0. The momentum range covered by the experiment of that paper corresponds to values of $k$ less than about $0.2a$, so the high momentum limit is not reached.

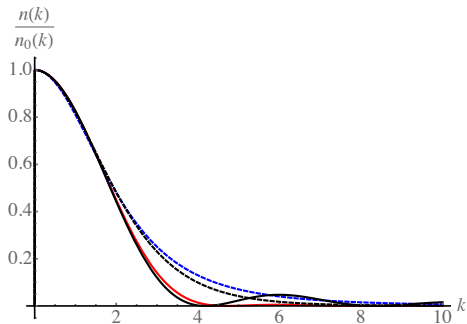

FIG. 3. (Color online) Comparison of $n(k)/n_0(k)$, for four different wave functions with $n(k) \equiv \widetilde{\phi}^2(k)$. Square well (red-solid), Hulthein (blue-dashed), surface delta (black-solid), and exponential (black-dashed) for $a/R = 10,000$. Dimensionless units are used (see text).

Our interest here is in understanding the regime of relatively low values of $k$ shown in Fig. 2. We may accurately use the relation $\sin x/x = 1 - x^2/6$ for values of $x$ up to unity. If $x = 1$ the left and right hand side of the previous equation differ by only 3%. This is sufficient for all of the values of $k$ appearing in Fig. 2. Then

$$T(k) \to T(0)(1 - \frac{1}{6} k^2 \langle r^2 \rangle), \tag{14}$$

with

$$\langle r^2 \rangle \equiv \frac{\int dr \, r^2 V(r) r\phi(r)}{\int dr \, V(r) r\phi(r)} \tag{15}$$

The results of computing $\langle r^2 \rangle$ for the different models are shown in Table I. The model-dependent values of $\langle r^2 \rangle$ given in Table I represent a violation of universality. The fall off with $k^2$ is greater for the square-well and surface delta models. Thus the behavior of $n(k)$ is qualitatively understood using Eq. (14):

$$n(k)/n_0(k) = (1 - \frac{1}{3} k^2 \langle r^2 \rangle), \tag{16}$$

which is valid (to within 6% or better) for values of $kr_e$ up to unity

The ratios $n(k)/n_0(k)$ for different models are shown in Fig 3. We that the models are indistinguishable for values of $k$ up to about $2/r_e$. For larger values of $k$ the fall-off is consistent with the quadratic behavior of the approximate result Eq. (16), with the rate of fall-off consistent with the results of Table I. The $H$ and $E$ models are very similar to each other, and so are $S$ and $SD$ models. But all of the differences show up only at $k > 2/r_e$. Thus the concept of universality can be extended to values above the limits given by Eq. (2) by including effects of the non-zero value of $r_e$. The different values of $\langle r^2 \rangle$ for a fixed value of $r_e$ shown in Table I represent a small violation of universality.

| Model | $a = 10000, r_e = 1$ |
|---|---|
| 0 | 0 |
| S | 0.46 |
| H | 0.22 |
| SD | 0.57 |
| E | 0.29 |

TABLE I. The quantity $\langle r^2 \rangle$ for the $0, S, H, SDE$ models. The ratio of $\langle r^2 \rangle$ to unity gives the quantity $\lambda$ of the text.

The results obtained here use a fixed scattering length. However, the value of the scattering length is important in and of itself. The results Eq. (14), Eq. (16) and Table I tell us that the density $n(k) = n_0(k)(1 - \lambda k^2 r_e^2)$ with $\lambda$ of the order of unity. Different models have different values of $\lambda$, even if the effective range of the different models is the same. Moreover, should $\lambda r_e^2$ be comparable to unity, the function $n(k)$ would have very little resemblance to $n_0(k)$.

We next explore the consequences that non-zero values of $\langle r^2 \rangle$ have for the energy decomposition [8]. For the two-body system at rest, and for large scattering lengths, this reads:

$$B = -\frac{1}{a^2} = \frac{4\pi}{(2\pi)^3} \int_0^\infty dk\, k^4 \left[n(k) - \frac{C}{k^4}\right] + \frac{C}{4\pi a}, \tag{17}$$

in our units, with the contact $C = 8\pi/a$. The energy relation of [13] is essentially the same as in Eq. (17) because an upper limit $\Lambda$ can be taken to be infinite by adjusting a bare coupling constant to reproduce a given value of $a$. Evaluating the integral for the zero range model gives a value of $-3/a^2$, so for that model the energy relation correctly becomes: $-1/a^2 = -3/a^2 + 2/a^2$. It is worthwhile to rewrite Eq. (17) for the particular case of the zero range model as

$$1 = -\frac{a^2}{6\pi^2} \int_0^\infty dk\, k^4 (n_0(k) - \frac{8\pi}{ak^4}), \tag{18}$$

with $n_0(k) = \frac{8\pi a^3}{(1+k^2 a^2)^2}$.

But for models with a non-zero effective range, with a general asymptotic behavior: $\widetilde{\phi}^2(k)$ falls at least as fast as $1/k^8$. This causes the integral appearing in the right-hand side of Eq. (17) to diverge linearly, and the energy relation of Eq. (17) cannot be valid as written. Indeed, Fourier transforming each of the wave functions, $\widetilde{\phi}(k)$, appearing in the quantity $\frac{4\pi}{(2\pi)^3} \int_0^\infty dk\, k^4\, \widetilde{\phi}(k)^2$ shows that it is the expectation value of the kinetic energy: $\langle \phi | (-\nabla^2) | \phi \rangle$, which is finite for models with a non-zero effective range. If one takes the upper limit of the integral in Eq. (17) to be a finite quantity $\Lambda$ the second term is $-C\Lambda$. This means that the relatively small differences between the densities $n(k)$ and $n_0(k)$ cause a linear divergence in the energy relation of Eq. (17), and cannot be neglected. This is the origin of our previous statement that the formula $C = \lim_{k\to\infty} k^4 n(k)$ cannot be taken literally. The value of infinity is too large. We seek to fix this problem

This problem can be fixed, and a version of the energy decomposition can be recovered. Let's choose the upper limit of the integral to be $\Lambda$ and define its ratio to the zero range result of $-3/a^2$ to be a function $R_\Lambda$, with

$$R(a, \Lambda) \equiv -\frac{a^2}{6\pi^2} \int_0^\Lambda dk\, k^4 (n(k) - \frac{8\pi}{ak^4}). \tag{19}$$

Note that if $n(k)$ is well approximated by $n_0(k)$ over the range of the integrand, then $R(a, \Lambda)$ would be close to unity and the sum rule of Eq. (17) would be corrected simply by replacing the infinite upper limit of the integral in Eq. (17) by $\Lambda$.

We analyze Eq. (19) by subtracting and adding Eq. (18), so that

$$R(a, \Lambda) = 1 - \frac{a^2}{6\pi^2} \int_0^\Lambda dk\, k^4(n(k) - \frac{8\pi}{ak^4}) + \frac{a^2}{6\pi^2} \int_0^\infty dk\, k^4(n_0(k) - \frac{8\pi}{ak^4}). \tag{20}$$

$$= 1 - \frac{a^2}{6\pi^2} \int_0^\Lambda dk\, k^4(n(k) - n_0(k)) + \frac{a^2}{6\pi^2} \int_\Lambda^\infty dk\, k^4(n_0(k) - \frac{8\pi}{ak^4}) \tag{21}$$

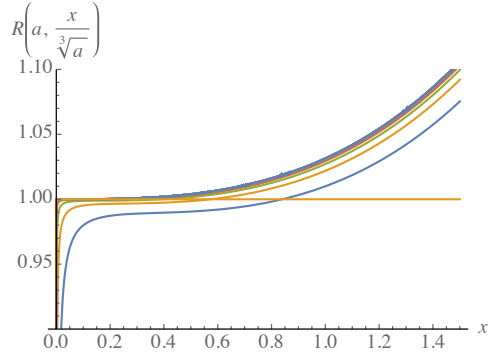

FIG. 4. (Color online) The function $R(a, \Lambda = x/a^{(1/3)})$ as a function of the variable $x$ for values of $a$ ranging from $10^4$ to $10^{19}$. The lowest curve represents $a = 10^4$. Dimensionless units are used (see text).

The second integral term of the above equation is expected to be very small. It may be evaluated as:

$$\frac{a^2}{6\pi^2} \int_\Lambda^\infty dk\, k^4 (n_0(k) - \frac{8\pi}{ak^4}) \approx \frac{-8}{3\pi\Lambda a} + \mathcal{O}(1/(\Lambda a)^3) \tag{22}$$

Note that $\Lambda a \gg 1$ because $a$ is large. Thus any terms falling as a power of $a$ can be safely neglected in the limit of large scattering length. The net result is that for that limit

$$R(a, \Lambda) = 1 - \frac{a^2}{6\pi^2} \int_0^\Lambda dk\, k^4 (n(k) - n_0(k)) \equiv 1 - \Delta R(a, \Lambda). \tag{23}$$

We shall see that the above expression can be approximated by unity for appropriate values of $\Lambda$. To see this use $n(k)/n_0(k) = (1 - \frac{1}{3}k^2 \langle r^2 \rangle)$, in its range of applicability ($kr_e < 1$), so that

$$\Delta R(a, \Lambda) = \frac{4a^5 \langle r^2 \rangle}{9\pi} \int_0^\Lambda dk\, \frac{k^6}{(1+k^2a^2)^2} \approx \frac{4}{27\pi} a \langle r^2 \rangle \Lambda^3. \tag{24}$$

The conventional approach to the upper limit $\Lambda$ in the integral appearing in Eq. (23) would be to take $\Lambda$ to be $x/r_e$, with $x$ on the order of 0.1. In that case, Eq. (23) tells us that $\Delta R(a, \Lambda)$ would grow linearly without bound as $a$ increases, and the energy sum rule would not be not valid.

Instead we derive the necessary criterion for validity by setting $\Delta R(a, \Lambda)$ to 0.1. This gives

$$\Lambda = \frac{1.28}{(a \langle r^2 \rangle)^{1/3}}, \tag{25}$$

and the value of the upper limit is smaller than $x/r_e$ for values of $a$ greater than about 1000. The quadratic approximation to the density, Eq. (16) remains valid for such values of $a$. The net result is that the energy decomposition can be written as

$$-\frac{1}{a^2} = \frac{4\pi}{(2\pi)^3} \int_0^\Lambda dk\, k^4 \left[n(k) - \frac{C}{k^4}\right] + \frac{C}{4\pi a}, \tag{26}$$

to an accuracy of at least 10%. The upper limit Eq. (25) is a surprise as many who would expect instead that $\Lambda \propto x/r_e$, with $x$ a small number in the vicinity of 0. To check that this is the correct upper limit we plot the function $R_{x/a^{1/3}}(a)$ (as obtained for the Hulthein potential) for a huge range of values of $a$ as a function of $a$ and $x$ in Fig. 4. If the dependence given by Eq. (24) were exact the different curves would coalesce into a single one, and this occurs to a large extent for $a > 10^5$. We also see that $R(a, \Lambda) \approx 1$ for a large range of values of $x$.

The calculations of $R(a, \Lambda)$ lead to the statement:

$$C \approx k^4 n(k),\, 0 < k < \Lambda, \tag{27}$$

*if one wishes to maintain the energy sum rule.*

The density-density correlation at short distances [8, 17] is discussed next. For a system of atoms with two different spins (1,2), using the zero-range model yields the result:

$$\langle n_1(\mathbf{R} + \tfrac{1}{2}\mathbf{r})\, n_2(\mathbf{R} - \tfrac{1}{2}\mathbf{r})\rangle \to \frac{1}{16\pi^2}\left(\frac{1}{r^2} - \frac{2}{ar}\right)\mathcal{C}(\mathbf{R}) \tag{28}$$

for small separations $r$, where $n_{1,2}$ are density operators. The function of $r$ arises from the singular part of the square of the wave function $\phi_0(r) \propto (1/r - 1/a)$. Integrating Eq. (28) both $\mathbf{r}_1, \mathbf{r}_2 = \mathbf{R} \pm \mathbf{r}/2$ over a ball of radius $s$, defining an operator

$$\mathcal{O}(s)f(s) \equiv \int d^3r_1 d^3r_2 \theta(s-r_1)\theta(s-r_2) f(s) \tag{29}$$

Refs. [8, 17] find

$$N_{\text{pair}}^{(0)}(\mathbf{R}, s) \equiv \mathcal{O}(s)\langle n_1(\mathbf{R}+\tfrac{1}{2}\mathbf{r})\, n_2(\mathbf{R}-\tfrac{1}{2}\mathbf{r})\rangle = \tfrac{s^4}{4}\mathcal{C}(\mathbf{R}). \tag{30}$$

The volume of the ball is $V = 4\pi s^3/3$. The $s^4$ behavior, which arises from the $1/r^2$ term in the square of $\phi(r)$, interesting because the number of pairs inside that ball scales as $V^{4/3}$ instead of the expected $V^2$ behavior [17].

Here we evaluate $N_{\text{pair}}(\mathbf{R}, s)$ using a general model of $\phi(r)$, which must be finite at the origin. Consider the quantity

$$N_{\text{pair}}(\mathbf{R}, s) \equiv \mathcal{O}(s)\langle n_1(\mathbf{R}+\tfrac{1}{2}\mathbf{r})\, n_2(\mathbf{R}-\tfrac{1}{2}\mathbf{r})\rangle = F(s)\mathcal{C}(\mathbf{R})$$
$$F(s) \equiv \mathcal{O}(s)\phi(|\mathbf{r}_1 - \mathbf{r}_2|))^2. \tag{31}$$

The function $F(s)$ may be evaluated in terms of $\widetilde{\phi}$, so that

$$F(s) = \int \frac{d^3k}{(2\pi)^3} \frac{d^3k'}{(2\pi)^3} \int d^3r_1 d^3r_2 \theta(s-r_1)\theta(s-r_2)$$
$$\times \widetilde{\phi}(k)\widetilde{\phi}^*(k')\exp[i(\mathbf{k}-\mathbf{k}')\cdot(\mathbf{r}_1-\mathbf{r}_2)]. \tag{32}$$

The integrals over $r_{1,2}$ may be evaluated to yield

$$F(s) = V^2 \int \frac{d^3k}{(2\pi)^3} \frac{d^3k'}{(2\pi)^3} \widetilde{\phi}(k)\widetilde{\phi}(k')\left(\frac{3j_1(qs)}{qs}\right)^2, \tag{33}$$

where $q \equiv |\mathbf{k}-\mathbf{k}'|$, and $j_1(x)$ is a spherical Bessel function of order unity, with $\lim_{x\to 0} j_1(x) = x/3$. These integrals are uniformly convergent because of the $1/k^4$ (or faster) asymptotic behavior of $\widetilde{\phi}(k)$. Thus, in seeking the behavior for small values of $s$ we may take the limit $qs \to 0$ inside the integral. This procedure leads to

$$\lim_{s\to 0} F(s) = V^2 \phi^2(0) \tag{34}$$

with the wave function at the origin $\phi(0)$ a finite quantity for all models that assume that the range of interaction is not zero. The use of an appropriate wave function. leads to a $V^2$ dependence.

Our results show that the influence of a non-zero effective range causes significant corrections to the zero-range model of the density $n_0(k)$, with the details depending on the value of the effective range $r_e$. The relation between $r_e$ and the size parameter of the each model depends on the given model, and can be regarded as a violation of universality. Given a fixed value of $r_e$, the density $n(k)$ is universal for values of $k$ up to about $2/r_e$. The corrections to $n_0(k)$ cause the energy decomposition to fail, and the pair density $N_{\text{pair}}(\mathbf{R}, s)$ to vary as $s^6$. The divergent energy decomposition of Eq. (17) is rescued by using a cutoff momentum that $\sim 1/(a^{1/3}\langle r^2\rangle^{1/3})$. Moreover, the contact $C$ cannot be thought of as the coefficient of the asymptotic $1/k^4$ behavior of the density. But Eq. (27) finds that this coefficient is valid for a finite range of momenta.

The violations of universality discussed here involve non-zero values of $r_e$. For a fixed value, these. violations are apparent only for $k > 2/r_e$. The influences of the effective range are important because they are related to the underlying detailed nature of the atom-atom interaction. Determining the details of $n(k)$ and its dependence on the effective range could reveal much about the underlying interactions.

This. work was supported by the U.S. Department of Energy Office of Science, Office of Nuclear Physics under Award No. DE-FG02-97ER-41014. I thank E. Braaten, A. Bulgac, A. Cherman and M.J. Savage for useful discussions.